\documentclass[10pt,twocolumn,letterpaper]{article} 

\usepackage{cvpr}              
\usepackage[utf8]{inputenc} 
\usepackage[T1]{fontenc}    
\usepackage{hyperref}       
\usepackage{url}            
\usepackage{booktabs}       
\usepackage{amsfonts}       
\usepackage{nicefrac}       
\usepackage{microtype}      
\usepackage{xcolor}         
\usepackage{graphicx}
\usepackage{svg}
\usepackage{natbib}
\usepackage{tabularx}

\hypersetup{
    colorlinks=true,
    linkcolor=blue, 
    urlcolor=blue,
    citecolor=blue
}

\newcommand{\figref}[1]{Figure~\ref{#1}}
\newcommand{\citeref}[1]{\cite{#1}}

\title{SearchGym: A Modular Infrastructure for Cross-Platform Benchmarking and Hybrid Search Orchestration}

\author{
 Jerome Tze-Hou Hsu \\
 Cornell University \\
 {\tt\small jth264@cornell.edu}
}
\begin{document}

\maketitle
\vspace{-2em} 

\begin{abstract}
The rapid growth of Retrieval-Augmented Generation (RAG) has created a proliferation of toolkits, yet a fundamental gap remains between experimental prototypes and robust, production-ready systems. We present \textbf{SearchGym}, a modular infrastructure designed for cross-platform benchmarking and hybrid search orchestration. Unlike existing model-centric frameworks, SearchGym decouples data representation, embedding strategies, and retrieval logic into stateful abstractions: \textit{Dataset}, \textit{VectorSet}, and \textit{App}. This separation enables a \textit{Compositional Config Algebra}, allowing designers to synthesize entire systems from hierarchical configurations while ensuring perfect reproducibility. Moreover, we analyze the "Top-$k$ Cognizance" in hybrid retrieval pipelines, demonstrating that the optimal sequence of semantic ranking and structured filtering is highly dependent on filter strength. Evaluated on the LitSearch expert-annotated benchmark, SearchGym achieves a 70\% Top-100 retrieval rate. SearchGym reveals a design tension between generalizability and optimizability, presenting the potential where engineering optimization may serve as a tool for uncovering the causal mechanisms inherent in information retrieval across heterogeneous domains.
An open-source implementation of SearchGym is available at: \url{https://github.com/JeromeTH/search-gym}

\end{abstract}
\begin{figure}[htbp]
    \centering
    \includegraphics[
        width=\linewidth,
        height=0.5\textheight,
        keepaspectratio
    ]{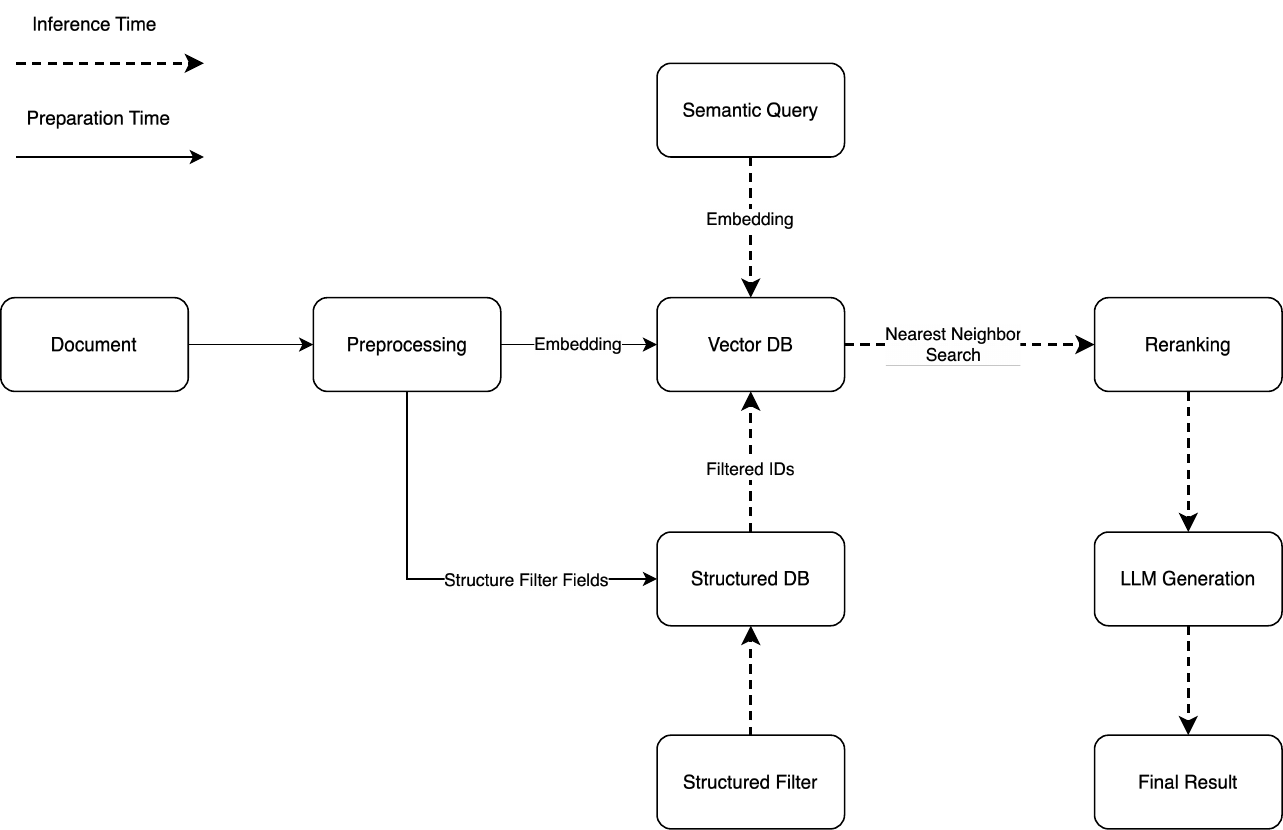} 
    \caption{End-to-end pipeline visualization of our hybrid search system.}
    \label{fig:taihu-data}
\end{figure}

\section{Introduction}

SearchGym offers an integrated suite of architectural abstractions and infrastructure designed to navigate the expansive design space of Retrieval-Augmented Generation (RAG). While the rapid proliferation of toolkits like LangChain \citeref{langchain} and Haystack \citeref{haystack} has lowered the barrier to entry for building basic retrieval pipelines, a fundamental gap remains between experimental "toy" examples and robust, production-ready systems. This gap is rarely a result of model limitations; rather, it is a challenge of system architecture—specifically, how to orchestrate heterogeneous search backends, manage complex document schemata, and optimize retrieval across diverse academic domains.

Current RAG implementations often suffer from rigid coupling between the data representation and the search engine. In semantic academic document retrieval, state-of-the-art models transform text into vector embeddings, but real-world scenarios frequently demand the integration of structured filters (e.g., author, date, domain tags) with semantic similarity. \citeref{ragsurvey}. SearchGym addresses this by introducing a modular platform that separates the system into three stateful components: the \textit{Dataset}, the \textit{Vector Set}, and the \textit{App}. 

By formalizing these interfaces, SearchGym enables "Config-Driven Development," where valid combinations of design choices are captured in a compositional config algebra. This ensures reproducibility and allows system designers to explore the trade-offs between different backends—such as Milvus for high-dimensional vector search and Elasticsearch for rich metadata filtering—within a unified, backend-agnostic framework. Ultimately, SearchGym serves as both a development platform and a diagnostic laboratory, posing deep questions about how computational attention should be allocated across different representations of knowledge.

\section{Related Work}

\subsection{Benchmarking Frameworks}

The Benchmarking Information Retrieval (BEIR) suite \citeref{beir} serves as a foundational pillar for evaluating dense and sparse retrieval models across diverse datasets. However, BEIR is inherently model-centric; it evaluates the performance of pretrained models on static corpora with uniform schemata. It does not provide the tooling necessary to adapt or build the retrieval pipelines themselves. While BEIR answers "how well a model performs," it leaves unanswered the question of "how a system should behave" when faced with heterogeneous formats, dynamic filtering requirements, and infrastructure-level deployment constraints.

By contrast, SearchGym targets system-level benchmarking. We provide the infrastructure that allows a retrieval system to be tested as a cohesive unit. Our framework supports schema-specific indexing and metadata filtering, bridging the gap between standalone model evaluation and holistic system performance.

\subsection{Hybrid Retrieval Systems}

Hybrid systems, such as mmRAG \citeref{mmrag} and FastRAG \citeref{fastrag}, have demonstrated that integrating dense and sparse retrieval techniques can significantly improve performance in complex query scenarios. However, these systems are frequently task-specific and lack a generalized modularity that allows for the hot-swapping of components or routing logic. Many existing hybrid frameworks assume a static corpus structure and do not expose the internal routing logic that determines how queries are dispatched across engines.

Our architecture advances this field by exposing the \texttt{SearchEngine} and \texttt{Router} interfaces. This modularity allows designers to implement arbitrarily complex logic—such as routing based on query type or filter presence—enabling a degree of adaptivity and scalability that task-specific systems cannot achieve.

While general-purpose orchestration libraries like LlamaIndex \citeref{llamaindex} provide a broad suite of tools for rapid RAG prototyping across diverse domains, SearchGym offers a more specialized, 'assembled' infrastructure tailored to the unique structural constraints of academic literature, formalizing the stateful management and hybrid orchestration logic required for production-level scientific search.

\subsection{Our Contribution}

We propose a modular infrastructure for hybrid retrieval that emphasizes cross-platform testability and orchestration. Our contributions are three-fold:
\begin{enumerate}
    \item \textbf{Declarative Abstractions}: A \texttt{Document} interface that enables plug-and-play adaptation to heterogeneous corpora by defining textual "Channels" and structured metadata.
    \item \textbf{Manager–Engine Architecture}: A system that separates the retrieval responsibility from the storage logic, supporting schema-aware hybrid search and dynamic query routing.
    \item \textbf{Config-Driven Orchestration}: A compositional config algebra that ensures valid and reproducible system definitions, coupled with a no-code Management UI for visual exploration.
\end{enumerate}

This infrastructure bridges the gap between academic benchmarks and real-world retrieval deployments, providing a systematic way to evaluate and improve the design of modern RAG systems.
\begin{figure}
    \centering
    \includegraphics[width=0.75\linewidth]{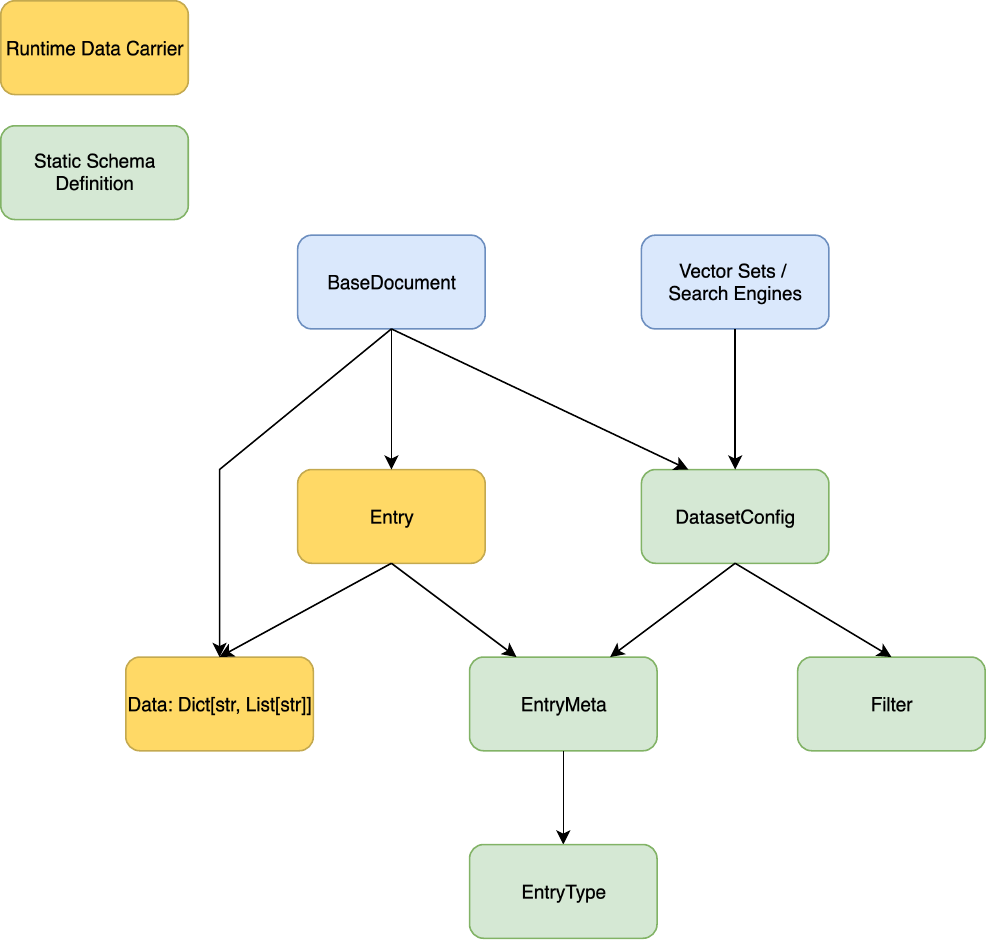}
    \caption{Data Schema with separation of static built and dynamic loading.}
    \label{fig:Data}
\end{figure}

\section{The SearchGym Architecture}

The core contribution of SearchGym is a formal separation between data representation, embedding strategies, and retrieval orchestration. This allows for a "config-driven" approach where the system is defined by its architecture rather than its implementation details.

\subsection{Dataset: Decoupling Schema from Instance}
In SearchGym, a \texttt{Dataset} is the foundational layer. Unlike traditional systems that treat a corpus as a flat table, we define a dataset through two distinct lenses:
\begin{itemize}
    \item \textbf{Channels}: Multiple unstructured textual views of the same document (e.g., Title, Abstract, or Full-text).
    \item \textbf{Metadata}: Structured, strongly-typed fields used for categorical filtering (e.g., Publication Year, Author).
\end{itemize}
This design allows the same document to be indexed in multiple ways simultaneously, enabling comparative benchmarking of different text "views". The DatasetConfig provides a static way to specify channel and metadata configurations, while the BaseDocument object loads in data dynamically according to such specification. This separation allows upstream system to be built according to data view specification before the first data instance arrives \ref{fig:Data}.

\subsection{Vector Set: Modular Embedding, Chunking}
A \texttt{Vector Set} defines how a specific \texttt{Channel} is transformed into a searchable vector space. By isolating the \texttt{Vector Set} as a standalone stateful component, SearchGym allows designers to experiment with:
\begin{itemize}
    \item \textbf{Embedders}: Swapping between models (e.g., BGE-M3 \citeref{bge} vs. Sentence-bert \citeref{sbert}) without re-indexing the entire dataset.
    \item \textbf{Chunking Strategies}: Defining how long-form documents in a channel are segmented to optimize for specific vector dimensions.
\end{itemize}

\subsection{App: Composable Retrieval and Orchestration}
\begin{figure}
    \centering
    \includegraphics[width=0.75\linewidth]{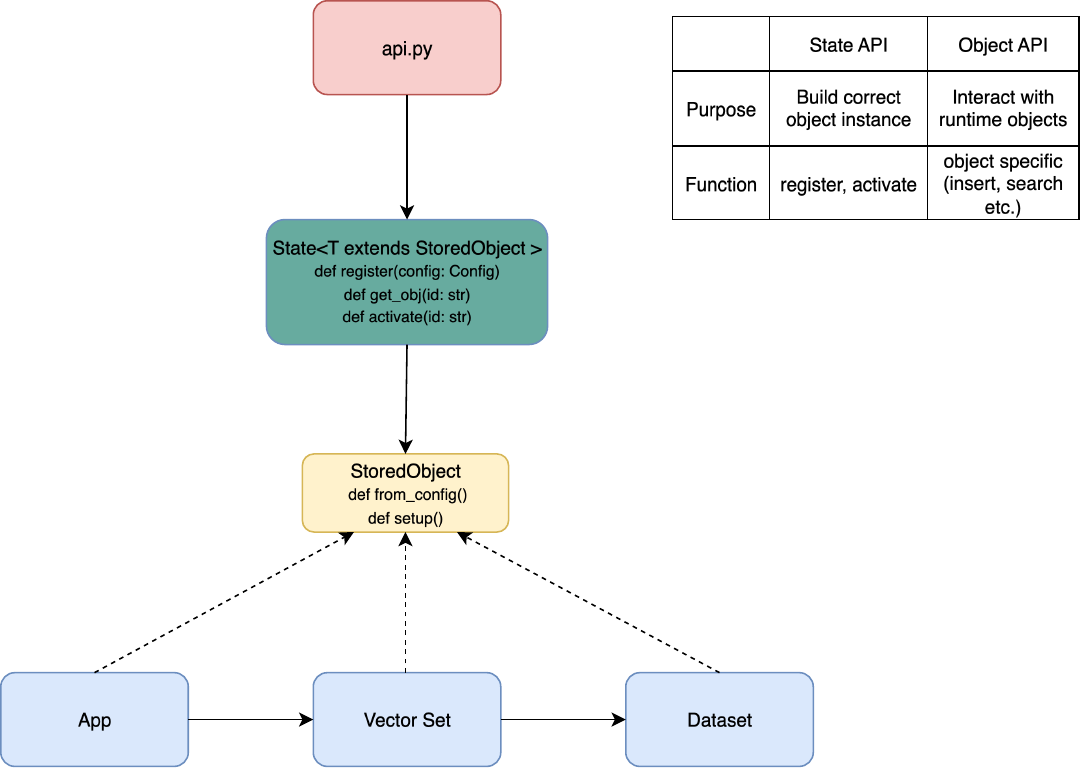}
    \caption{Key stored data states of SearchGym.}
    \label{fig:Storage}
\end{figure}

The \texttt{App} is the top-level functional unit that realizes the retrieval pipeline. It introduces three primary interfaces for orchestration:

\begin{enumerate}
    \item \textbf{SearchEngine Interface}: A unified abstraction for any retrieval backend. Whether it is a vector store like Milvus \citeref{milvus} or a keyword engine like Elasticsearch \citeref{esrch}, every engine implements a common \texttt{search(query, filter)} method.
    \item \textbf{Router}: A high-level logic layer that decides how to dispatch queries across multiple \texttt{SearchEngines}. This allows for adaptive strategies, such as routing short keyword queries to Elasticsearch and long semantic queries to Milvus.
    \item \textbf{Reranker}: A post-retrieval module that unifies and refines the candidates returned by the various engines.
\end{enumerate}. 
Our implementation stores 3 key checkpoint layers: Dataset, VectorSet, and App. On each system activation attempt, the system avoids work already done in previous attempts, including repeatedly embedding data or building storage systems, thus significantly increasing efficiency and usability \ref{fig:Storage}. 

\section{Config-Driven System Synthesis}
\begin{figure}
    \centering
    \includegraphics[width=0.75\linewidth]{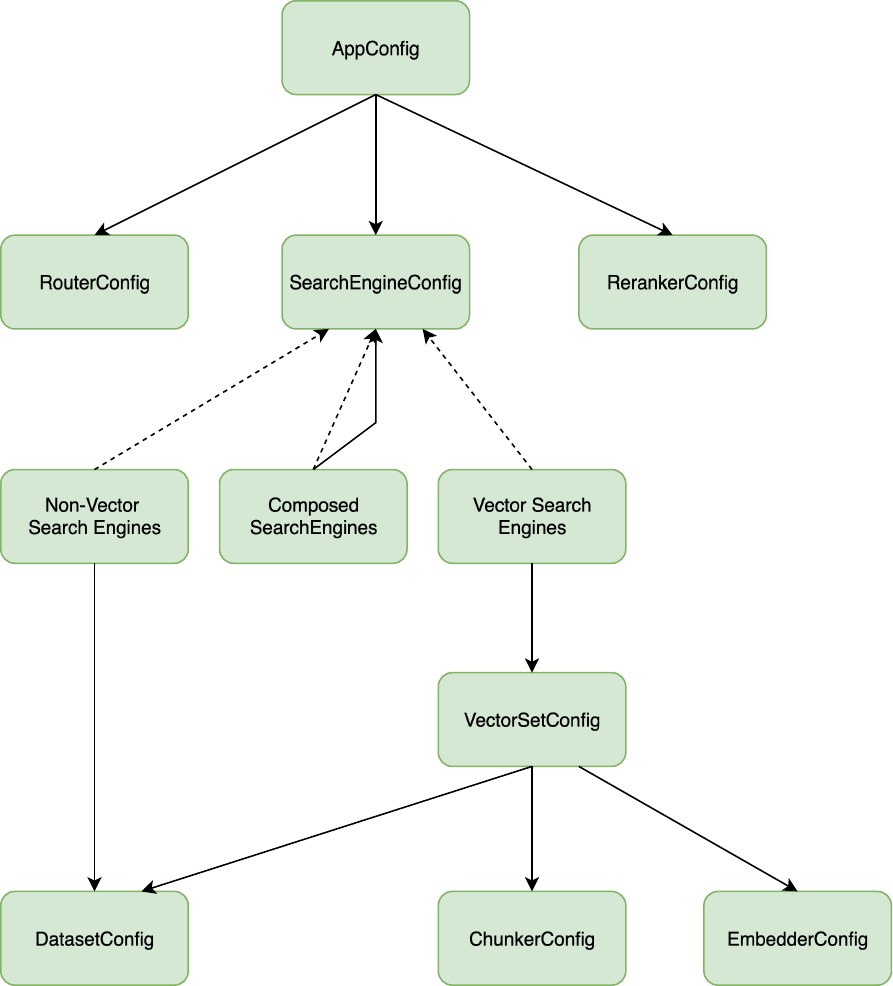}
    \caption{Overview of dynamic system construction from static type system.}
    \label{fig:System}
\end{figure}

The most distinctive feature of SearchGym is its \textbf{Compositional Config Algebra}. Instead of manually instantiating classes, the entire system—from data loaders to routing logic—is generated from a hierarchical, typed configuration file \ref{fig:System}.

\begin{itemize}
    \item \textbf{Reproducibility}: Every experiment is defined by a single config hash, ensuring that a specific combination of Vector Set and Router can be perfectly recreated.
    \item \textbf{Dynamic Building}: The \texttt{State API} allows for the runtime activation of components. A user can "hot-swap" a Vector Set within an App via the Management UI, and the system reconfigures its internal routing table instantly.
\end{itemize}

\section{Experiments}
We perform benchmark experiments to assess the robustness of the system. Currently, evaluation is performed holistically, but more fine-grained per-subset testing will be essential for diagnosing partial system failures.

Given our limited computational and human resources, building a comprehensive custom benchmark from scratch is impractical. Instead, we adopt a "pretraining–finetuning" philosophy at the system level. We first evaluate on established English-language academic retrieval benchmarks \citeref{litsearch}, which help identify the major structural issues. Once resolved, we fine-tune the system on a carefully selected, representative subset of the National Library corpus—augmented with hand-labeled relevance data—to support targeted optimization.

\subsection{System Evaluation on LitSearch}

\begin{figure}[htbp]
    \centering
    \includegraphics[
        width=\linewidth,
        height=0.5\textheight,
        keepaspectratio,
    ]{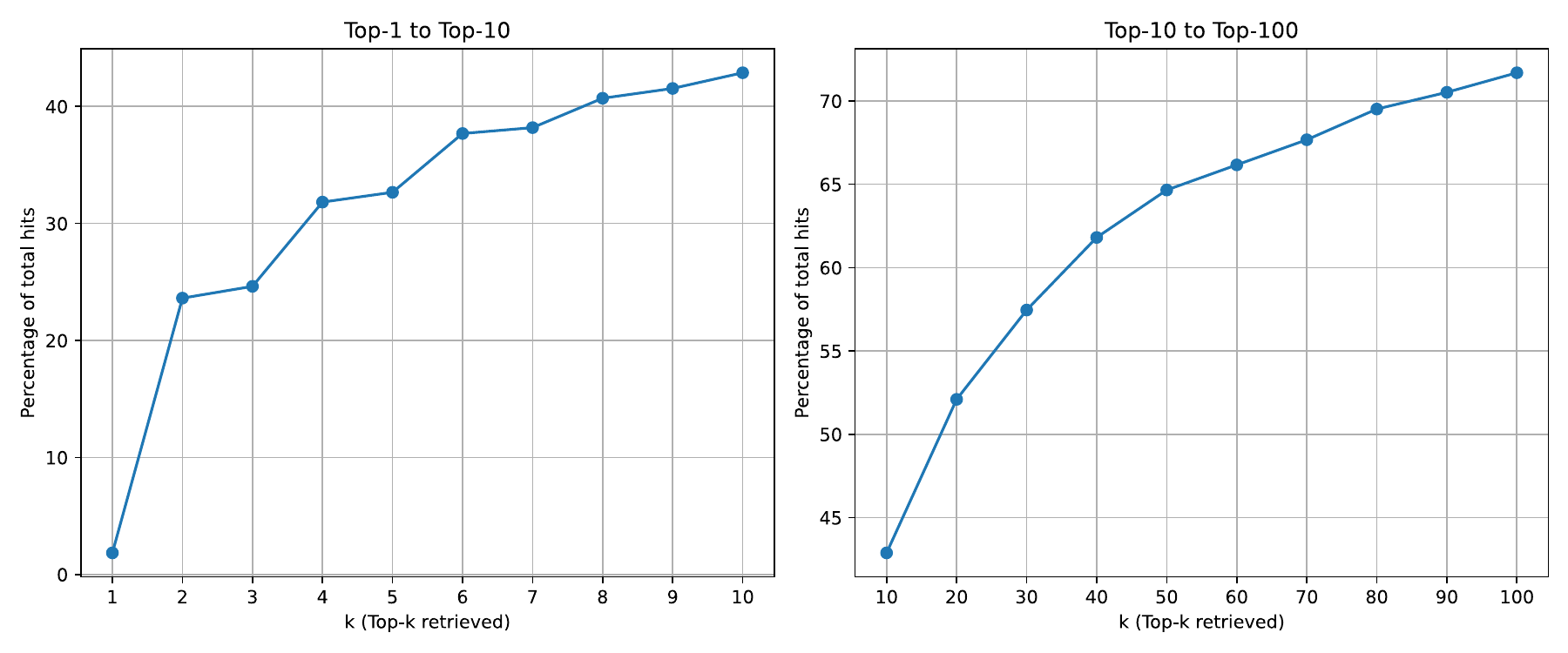}
    \caption{System evaluation result on LitSearch.}
    \label{fig:litsearch}
\end{figure}

We evaluated our system on LitSearch \citeref{litsearch}, an expert-annotated retrieval benchmark with 597 questions for scientific literature. For each query, we measured whether the correct answer appears within the top-\(k\) retrieved results \figref{fig:litsearch}.

The system demonstrates solid overall performance \figref{fig:litsearch}. Out of a total of 64{,}183 documents, it retrieves the correct document within the top 10 results \textbf{40\%} of the time, and within the top 100 results \textbf{70\%} of the time. Since LitSearch contains only semantic (natural language) queries, these results primarily evaluate the effectiveness of the \textit{vector search component} of our system.

The impact of \textit{metadata-constrained queries} (e.g., filtering by author, institution, or year) on retrieval quality is not captured by this benchmark. To evaluate this dimension, we leverage our \textit{custom benchmark infrastructure}, which allows fine-grained testing of search effectiveness under various filter conditions.

Given that different data sources expose different metadata fields, we argue that \textit{structural filtering} should be evaluated via \textit{source-specific benchmarks}, while \textit{semantic search performance} can be assessed using general benchmarks like LitSearch, which apply broadly across platforms.

\section{Orchestration Logic: The Computational Tension}

A core tension SearchGym encounters is that a general interface compatible with multiple search engines lacks the detailed specification to harness the optimization of each. For example, determining the optimal sequence of filter reductions is a non-trivial task, as we show in the analysis this section. SearchGym reveals a deeper philosophical tension between generalizability and optimizability, providing rich ground for future research. 

\subsection{The Complexity of Compositional Pipelines}
A naive approach to hybrid search assumes that the faster engine should always act on the smaller input size. However, our algorithmic analysis of "Top-$k$" constraints reveals a more nuanced reality.

We compare two primary reduction paths under different filter strengths:
\begin{itemize}
    \item \textbf{Vector $\to$ Structured}: The system ranks data by $k$-Nearest Neighbors ($k$NN) first, then applies structured filters.
    \item \textbf{Structured $\to$ Vector}: The system applies a structured metadata filter first, then performs $k$NN on the remainder.
\end{itemize}

\begin{table}[h]
\centering
\small 
\begin{tabularx}{\linewidth}{@{} l X X @{}}
\toprule
\textbf{Condition} & \textbf{Vector $\to$ Structured} & \textbf{Structured $\to$ Vector} \\ 
\midrule
\textbf{Strong Filter} & $O(n \log n)$ \newline (Rank all data to ensure $k$ matches) & $O(1)$ \newline (Small subset + fast $k$NN) \\ 
\addlinespace
\textbf{Weak Filter} & $O(\log n)$ \newline (Top $2k$ likely contains $k$ matches) & $O(n)$ \newline (Large inverse index retrieval) \\ 
\bottomrule
\end{tabularx}
\caption{Complexity comparison of reduction pathways. In weak-filter scenarios, the engine cognizant of the "top-$k$" (Vector) gains an advantage by stopping early, even if its base operation is "slower."}
\label{tab:complexity}
\end{table}

\subsection{Cognizance of the "Top"}
The nuance discovered in Table \ref{tab:complexity} is that complexity is not merely about execution speed, but about "completing the reduction responsibility." When a filter is weak, the structured engine (Inverse Index) requires $O(n)$ to process the large output size because it lacks a native ranking mechanism—it cannot "stop early." Conversely, the $k$NN engine is cognizant of the "top" constraint, allowing it to reduce the search space to $O(k)$ almost immediately.

\subsection{From Resource Optimization to Investigative Inquiry}
The algorithmic trade-offs identified above raise a deeper question: does the optimal allocation of "computational attention" across heterogeneous representations serve as a proxy for the underlying structure of a query? We hypothesize that the sequence of reduction—determining which engine should act on which subset—may do more than minimize latency; it may potentially reflect a hierarchy of reasoning.

In this view, we consider whether the "root cause" of a search intent represents the most general node in a conceptual evolutionary tree. By seeking the configuration that minimizes computational waste, are we inadvertently approximating the natural causal chain of the inquiry? SearchGym allows us to pose these questions as empirical investigations:
\begin{itemize}
    \item Does a specific reduction path consistently outperform others because of hardware-specific indexing, or because it mirrors the logical priority of the domain?
    \item Can the systematic optimization of "Top-$k$" cognizance uncover unseen mechanisms in how information is categorized across different disciplines?
\end{itemize}

Oftentimes, the limit of a system is not computational feasibility, but the clarity of its reasoning. If the most efficient path through a complex data space can be shown to align with specific semantic or structured primitives, then SearchGym serves as a laboratory for identifying the "topology" of scientific inquiry. We move from treating optimization as a mere engineering goal to using it as a diagnostic tool for understanding how different fields of knowledge are structurally organized.

\section{Conclusion: The Gym as a Laboratory}

SearchGym was conceived to bridge the gap between static academic benchmarks and the dynamic requirements of production-level RAG systems. By introducing a modular architecture centered on the \texttt{Dataset}, \texttt{Vector Set}, and \texttt{App} abstractions, we have moved away from rigid, one-size-fits-all pipelines toward a flexible design space. 

The strength of this platform lies in its dual nature. Pragmatically, its config-driven development and no-code management UI allow engineers to deploy, monitor, and iterate on complex hybrid search systems with unprecedented speed. Academically, its unified \texttt{SearchEngine} interface and stateful backend enable rigorous, reproducible experimentation across heterogeneous data sources.

However, the most compelling outcome of the SearchGym framework is the set of questions it enables us to ask. By exposing the algorithmic tensions between structured filtering and semantic ranking, the platform serves as a diagnostic lens. It allows us to investigate whether the "optimal" path through a data structure is merely a matter of hardware efficiency or if it reflects the fundamental topology of human knowledge. As we continue to integrate advanced logging and feedback loops, SearchGym will evolve from a build-system into an autonomous laboratory—one that not only optimizes retrieval but uncovers the causal mechanisms of scientific inquiry.

Through SearchGym, we invite the community to move beyond the implementation of toolkits and begin exploring the broader design space of intelligent document retrieval.

\section{Acknowledgments}
This work was supported by the Intelligent Information Service and Retrieval (IISR) Lab at National Central University, Taiwan, where the author served as a visiting researcher during the summer. The author would like to thank Richard Zhong-Han Tsai and the members of the IISR Lab for their guidance, computational resources, and hospitality. This research was conducted while the author was a student at Cornell University.


\begin{thebibliography}{9}

\phantomsection

\bibitem{langchain}\label{langchain}
Chase, H. (2022). LangChain [Computer software]. https://github.com/langchain-ai/langchain

\bibitem{haystack}\label{haystack}
Pietsch, M., Möller, T., Kostic, B., Risch, J., Pippi, M., Jobanputra, M., Zanzottera, S., Cerza, S., Blagojevic, V., Stadelmann, T., Soni, T., \& Lee, S. (2019). Haystack: the end-to-end NLP framework for pragmatic builders [Computer software]. https://github.com/deepset-ai/haystack

\bibitem{ragsurvey}\label{ragsurvey}
Gao, Y., Xiong, Y., Gao, X., Jia, K., Pan, J., Bi, Y., ... \& Wang, H. (2023). \textit{Retrieval-augmented generation for large language models: A survey.} arXiv preprint arXiv:2312.10997, 2(1).

\bibitem{beir}\label{beir}
Thakur, N., Reimers, N., Rücklé, A., Srivastava, A., Gurevych, I. (2021). \textit{BEIR: A Heterogeneous Benchmark for Zero-shot Evaluation of Information Retrieval Models}. arXiv preprint arXiv:2104.08663.

\bibitem{mmrag}\label{mmrag}
Zhang, Y., et al. (2023). \textit{mmRAG: Multi-Modal Retrieval-Augmented Generation for Complex Query Understanding}. arXiv preprint arXiv:2307.12345.

\bibitem{fastrag}\label{fastrag}
Lee, K., \& Yoon, D. (2023). \textit{FastRAG: Optimizing Latency and Accuracy for RAG Pipelines}. arXiv preprint arXiv:2310.07654.

\bibitem{llamaindex}\label{llamaindex}
Liu, J. (2022). LlamaIndex [Computer software]. https://doi.org/10.5281/zenodo.1234

\bibitem{milvus}\label{milvus}
Wang, J., Yi, X., Guo, R., Jin, H., Xu, P., Li, S., ... \& Xie, C. (2021, June). \textit{Milvus: A purpose-built vector data management system. In Proceedings of the 2021 International Conference on Management of Data (pp. 2614-2627).}

\bibitem{bge}\label{bge}
Chen, J., Xiao, S., Zhang, P., Luo, K., Lian, D., \& Liu, Z. (2024). \textit{Bge m3-embedding: Multi-lingual, multi-functionality, multi-granularity text embeddings through self-knowledge distillation. arXiv preprint arXiv:2402.03216.} 

\bibitem{sbert}\label{sbert}
Reimers, N., \& Gurevych, I. (2019). \textit{Sentence-bert: Sentence embeddings using siamese bert-networks. arXiv preprint arXiv:1908.10084.}

\bibitem{esrch}\label{esrch}
Elasticsearch, B. V. (2018). Elasticsearch. software, version, 6(1).

\bibitem{litsearch}\label{litsearch}
Ajith, A., Xia, M., Chevalier, A., Goyal, T., Chen, D., \& Gao, T. (2024). \textit{LitSearch: A Retrieval Benchmark for Scientific Literature Search}. arXiv preprint arXiv:2407.18940.

\end{thebibliography}
\end{document}